\documentclass[twocolumn,showpacs,floats,amsmath,amssymb,nofootinbib]{revtex4}
\usepackage{graphicx}
\usepackage{dcolumn}
\usepackage{bm}
\usepackage{amsmath}
\usepackage{float}
\usepackage{color}
\usepackage{amsfonts}
\usepackage{amssymb}
\usepackage[all]{xy}
\begin{document}
\title{A thermodynamics revision of Rastall gravity}

\author{Miguel Cruz$^{a,}$\footnote{E-mail: miguelcruz02@uv.mx}, Samuel Lepe$^{b,}$\footnote{E-mail: samuel.lepe@pucv.cl} and Gerardo Morales-Navarrete$^{a,}$\footnote{E-mail: moralesnavarretegerardo@gmail.com}}
\affiliation{$^{a}$Facultad de F\'\i sica, Universidad Veracruzana 91000, Xalapa, Veracruz, M\'exico \\ $^{b}$Instituto de F\'{\i}sica, Facultad de Ciencias, Pontificia Universidad Cat\'olica de Valpara\'\i so, Avenida Brasil 2950, Valpara\'\i so, Chile\\}

\date{\today}

\begin{abstract}
In this work we study some aspects of the Rastall gravity, being the thermodynamics consistency of the model the core of this paper, for this purpose we will consider the dynamical equations of Rastall model in a flat FLRW geometry. Under a holographic description can be seen that this scenario for gravity contributes to the energy density of the fluid with an extra term that can be related to the deceleration parameter, providing a way to estimate the value of the Rastall parameter, termed as $\xi \lambda$, at present time. By adopting a specific Ansatz for the $\xi \lambda$ term it is possible to determine that the behaviour of the Hubble parameter in Rastall gravity has a similar aspect to the $\Lambda$CDM model at late times, but at thermodynamics level differs from the standard cosmology since the adiabatic behaviour for the entropy depends on the value of the parameter state, $\omega$. However, the entropy has a positive growth and simultaneously its convexity condition can be guaranteed; when other contributions are considered such as matter production and chemical potential, the adiabatic expansion can not be achieved, but the theory keeps its thermodynamics consistency. The chemical potential seems to have an interesting role since at effective level we could have a cosmological constant or phantom expansion in the model.    
\end{abstract}


\maketitle

%%%%%%%%%%%%%%%
\section{Introduction}
\label{intro}
%%%%%%%%%%%%%%%

In Ref. \cite{rastall1} Rastall proposed a modification to the theory of gravity in order to be compatible with astronomical observations, such theory carry a variable quantity that can play the role of a cosmological constant when it takes a specific value and can be adjusted to be consistent with Dicke's measurement of solar oblateness. On the other hand, Rastall himself in Ref. \cite{rastall} found a theory that possesses a varying parameter, commonly named Rastall parameter; this latter characteristic of his theory was obtained by assuming a non conservation condition for the energy-momentum tensor; in the limit in which the Rastall parameter tends to zero the Einstein theory is recovered. In general, the Rastall parameter it is elected conveniently by adopting an Ansatz philosophy.\\ 

Although the conservation of the energy-momentum tensor is one of the cornerstones of general relativity, it has recently been shown that theories in which this conservation condition is not fulfilled can be in agreement with the observational data; as first example of non conservative model we have the inclusion of diffusive effects in matter into the Einstein equations, as discussed in Refs. \cite{diff1, diff2}, the aforementioned effects could have a relevant role in the formation of large scales structures and also could provide a viable theoretical framework for the interacting scheme between the dark matter and dark energy \cite{pavon}.\\ 
Other non conservative model is discussed in Ref. \cite{minazolli}, where under the consideration of a general gravity/matter coupling, it was found that the energy transference between gravity and matter modifies the matter fluid current conservation and this depends uniquely on the nature of the matter fields. Alternative examples of non conservative models are discussed in Refs. \cite{harko1, harko2}, where some extensions of the $f(R)$ gravity given as $f(R,L_{m})$ or $f(R,T)$\footnote{$L_{m}$ denotes the matter Lagrangian and $T$ the trace of the energy-momentum tensor.}, provide new scenarios since the movement of test particles is generally non-geodesic due to the presence of an extra force acting on them. The non conservative model given by the introduction of the extra term $f(T_{\mu \nu}T^{\mu \nu}, L_{m})$ in the Einstein-Hilbert action was discussed in \cite{non1}. At cosmological level it was found that under this description the baryonic matter couples to spacetime as in General Relativity while for the dark sector (dark matter/dark energy) it is different, this model also provides an unifying scenario for dark matter and dark energy with no need of extra scalar fields as done in other works. In Ref. \cite{non2} it was discussed the violation of the conservation condition for the energy-momentum tensor in the Brans-Dicke model under a conformal transformation, this leads to the absence of vacuum solutions for the Einstein's equations in the Einstein frame. However, the physical viability of the Einstein and Jordan frames is still under investigation. Finally, an interesting discussion can be found in \cite{sudarsky}, where was stated that the discreteness of the spacetime at Planck scale (expected to appear in the quantum gravity formulation) it is manifested by the non conservation of the energy-momentum tensor of matter fields. Besides, also was found that the discrete nature of the spacetime could lead to the emergence of a dark energy term in the Einstein's equations.\\ 

Nowadays it is well accepted that the inclusion of modifications into the Einstein gravity or an extension of this one are necessary to describe some phenomenological aspects of the universe that have been revealed by the current observations, being the most challenging the dark energy problem \cite{liddle, adam}. Although the $\Lambda$CDM model seems to be sufficient to describe the current state of the universe, it has some inconsistencies such as the value of the cosmological constant \cite{weinberg} or it is not sufficient to explain the abundance of certain elements in nature, in this sense, it should be noted that within the scenario of standard cosmology, some elementary aspects of an expanding universe, such as its thermodynamics, are not yet fully understood when the final fate of the universe relies on a future singularity \cite{lima1, lima2, saridakis}, i.e., a phantom universe. However, when the standard scheme is complemented by the inclusion of dissipative effects in the cosmological fluid and all processes are considered as irreversible, then the resulting phantom universe is consistent at thermodynamics level \cite{odintsovdiss}, therefore it is important to explore the thermodynamics conditions of any cosmological model that admits a phantom regime.\\ 

Some results claim that the Rastall gravity is equivalent to Einstein theory, therefore Visser in his work concludes that the Rastall model does not provide any new scenario in the gravity description \cite{visser}. However, in Ref. \cite{darabi} it was shown that Rastall model is in fact an extended theory of gravity that could be helpful to solve some problems in observational cosmology and quantum gravity. So, adopting the latter point of view it is valid to question the consistency or physical viability of this model. At cosmological level was found with the use of observational data in Refs. \cite{cosmo1, cosmo2, cosmo3}, that the Rastall model provides an interesting scenario in the formation of large structures, but also was stated that in the Rastall cosmology the vacuum energy agglomerates, therefore the Rastall and $\Lambda$CDM cosmologies can be distinguishable only at the non linear regime of the evolution of cosmic perturbations. Besides, within the framework of Rastall gravity, in Ref. \cite{stars} was discussed the equilibrium of neutron stars. This study revealed that the Rastall scenario leads to deviations from stellar models studied in the scheme of general relativity, but the results are still compatible with observations.\\

In this work we focus on the thermodynamics compatibility of the Rastall model. As we will see, the model it is consistent since the second law can be guaranteed together with the convexity condition for the entropy, but also we can observe that differs from the standard description of a cosmological fluid since the adiabatic expansion for the universe is obtained for a single value of the parameter state, $\omega$. Besides, when matter production and chemical potential are included, the adiabatic expansion can not be recovered, however the model keeps consistent at thermodynamics level. It is worthy to mention that the value of chemical potential has influence on the value of the parameter state, at least at effective level. According to some conditions the effective parameter state could describe a cosmological constant type evolution or a phantom fluid, this opens a new cosmological scenario for the Rastall model. In Ref. \cite{epjc1} some thermodynamics aspects of the Rastall model at cosmological level were discussed. Also, in Ref. \cite{epjc2} by the introduction of some different kinds of corrections to the entropy, the validity of the first and second law of thermodynamics was tested.\\ 

Globally speaking and as we will see in this work, the thermodynamics description could also be a helpful tool to construct an adequate Rastall parameter since in the literature can not be found a strong physical criterion for its construction/election.\\
 
This work is organized as follows: In Sect. \ref{sec:basics}, we write some general aspects of Rastall gravity and the dynamics of the model it is shown for a FLRW geometry. By considering a holographic approach, we estimate some values for the Rastall parameter, $\lambda$, at present time. We also show that a value for the Rastall parameter can be found by constructing a density parameter for the {\it new} contribution of the Rastall model in the energy density, $\rho$ or by focussing on the matter sector, as we will see, the set of values obtained for the Rastall parameter from different points of view are consistent. In Sect. \ref{sec:hubble} we give an expression for the normalized Hubble parameter by integrating the continuity equation of the Rastall model, we perform a comparison with the $\Lambda$CDM model, we also discuss that in general to obtain explicit values for the cosmological quantities we depend on the election of the $\xi \lambda$ term. We comment some conditions that the $\xi \lambda$ term must satisfy if the phantom behaviour is desired. In Sect. \ref{sec:second} we discuss the conditions that must by satisfied by the Rastall gravity in order to obey the second law of thermodynamics and by the inspecting the conditions on the entropy $dS/dt > 0$ and $d^{2}S/dt^{2} < 0$, we show that the consistency at thermodynamics level can be guaranteed. In Sect. \ref{sec:creation} we provide the production/annihilation rate of matter in the Rastall gravity given in terms of the deceleration parameter, we also show that once the matter production effects are introduced in the model, the entropy is not longer a constant value. In Sect. \ref{sec:chemical} we extend the thermodynamics discussion of the model by introducing the chemical potential. In Sect. \ref{sec:elfinal} we write the conclusions and some perspectives of our work. 

%%%%%%%%%%%%%%%%%%%
\section{Dynamics of Rastall gravity}
\label{sec:basics}
%%%%%%%%%%%%%%%%%%%

In Ref. \cite{rastall} Rastall questioned the energy-momentum tensor conservation in a curved spacetime and he found that the following divergence condition 
\begin{equation}
\nabla^{\mu}T_{\mu \nu} = \lambda \nabla^{\mu}(g_{\mu \nu}R),
\label{eq:condition}
\end{equation}
leads consistently to the Einstein equations when $\lambda \rightarrow 0$, where $\lambda$ it is usually known as Rastall parameter and $R$ is the Ricci scalar. As he also stated in his work, the Rastall and Einstein gravities coincide in an empty spacetime, but differ one from the other in presence of matter. In Ref. \cite{lind} the rate of deviation between the predictions of a non conservative theory as Rastall gravity and general relativity was estimated using gaseous helium data.\\ 

Therefore, assuming the condition given in (\ref{eq:condition}), the Rastall equation of motion takes the form
\begin{equation}
G_{\mu \nu} + \xi \lambda g_{\mu \nu }R = \xi T_{\mu \nu},
\label{eq:rastall}
\end{equation}  
being $\xi$ an appropriate constant. It is important to point out that there is not an action formulation from which the previous equation of motion could be obtained. However, the Rastall model has been widely studied from a phenomenological point of view \cite{visser}. Despite this represents the main drawback of the model, one alternative to solve this fact could be given by the introduction of an auxiliary (external) field in the Einstein-Hilbert action by means of a Lagrange multiplier. This would provide two possibilities: the Rastall gravity could be naturally seen as an extension of the Einstein model, secondly, one could also assume that this new contribution to the Einstein gravity arises from the quantum effects described by the Rastall equation \cite{strings}.\\ 

Notice that condition (\ref{eq:condition}) is recovered by means of the Bianchi identity satisfied by the Einstein tensor in the equation of motion (\ref{eq:rastall}). By taking the trace of the previous equation one gets
\begin{equation}
R = \frac{\xi T}{4\xi \lambda -1},
\end{equation}
which leads to the following expression
\begin{equation}
G_{\mu \nu} = \xi \mathcal{T}_{\mu \nu},
\end{equation}
where we have defined
\begin{equation}
\mathcal{T}_{\mu \nu} := T_{\mu \nu} + \frac{\xi \lambda T}{1-4 \xi \lambda}g_{\mu \nu},
\end{equation}
then the previous expression must be conserved, i.e., its divergence must be zero. Also, the condition given by $\lambda \neq 1/(4 \xi)$, must be taken into account. On the other hand, if we consider the Eq. (\ref{eq:rastall}) and take its trace, it is worthy to mention that in an empty space we have $(4\xi \lambda - 1)R = 0$, therefore a twofold condition can be obtained for the equation of motion: (a) if $\lambda \neq 1/(4 \xi)$ we have that $G_{\mu \nu} = 0$, and (b) if $\lambda = 1/(4 \xi)$ implies that $G_{\mu \nu} = \mathcal{A}g_{\mu \nu}$, where $\mathcal{A}$ is a constant. As can be seen from the previous conditions, depending on the value of the Rastall parameter, we have the Einstein vacuum equation of motion or the dynamics of the Einstein gravity plus a cosmological constant. From this result the Rastall parameter it is generally interpreted as a varying cosmological {\it constant}. If we consider the flat FLRW line element, the modified Friedmann equation can be written as \cite{epjc1, epjc2}
\begin{equation}
3H^{2} = \xi (\rho + \rho_{R}),
\label{eq:fried}
\end{equation}  
and the acceleration equation
\begin{equation}
2\dot{H} = -\xi(\rho + p + \rho_{R} + p_{R}) = -\xi (1+\omega)\rho,
\label{eq:accel}
\end{equation}
where the dot it means derivative with respect to the cosmic time and $\rho$, $p$ are the energy density and pressure of the fluid, respectively. In this case the quantities $\rho_{R}$ and $p_{R}$ are the new contribution coming from the Rastall model into the energy density and pressure of the fluid. For this case it is obtained that $p_{R} = -\rho_{R}$, i.e., the Rastall theory contributes negatively to the pressure of the fluid, then we have a dynamical cosmological {\it constant}, as commented above. Note that in Eq. (\ref{eq:accel}) we have considered a barotropic equation of state for the matter sector, $p = \omega \rho$, and explicitly we have $\rho_{R} = 6\lambda (\dot{H}+2H^{2})$. Additionally, the continuity equation has the form
\begin{equation}
\frac{d}{dt}\left(\frac{\rho + 6\lambda \dot{H}}{1-4\xi \lambda}\right)+3H(1+\omega)\rho = 0,
\end{equation}  
if we insert the Eq. (\ref{eq:accel}) in the previous equation one gets
\begin{equation}
\frac{d}{dt}\left[\left\lbrace\frac{1 - 3 \xi \lambda (1+\omega)}{1-4\xi \lambda}\right\rbrace \rho\right]+3H(1+\omega)\rho = 0,
\label{eq:continuity}
\end{equation}
then, based on the fact that the energy density is in general a function of the redshift parameter, $z$, (or the scale factor by means of the relation, $1+z = a_{0}a^{-1}$), the factor of the density $\rho$  within the square brackets of the above equation can be also expressed as a function of the redshift through the Rastall parameter if we consider $\lambda = \lambda(z)$, yielding
\begin{equation}
\Delta(z) := \frac{3(1+\omega)}{4}\left(\frac{\xi \lambda(z)-1/3(1+\omega)}{\xi \lambda(z)-1/4}\right).
\label{eq:delta}
\end{equation}
Another way to see the that the Rastall parameter is in general a function of the redshift is the following, we can insert the Eq. (\ref{eq:accel}) in the given definition of $\rho_{R}$ and the resulting expression can be substituted in the modified Friedmann equation (\ref{eq:fried}), yielding \cite{epjc2}
\begin{subequations}
\begin{eqnarray}
\xi \lambda(z) &=& \frac{3H^{2}(z)-\xi \rho(z)}{3(4H^{2}(z)-\xi (1+\omega)\rho(z))},\label{eq:xilambda1} \\ 
&=& \frac{\xi \rho_{R}(z)}{3(4H^{2}(z)-\xi (1+\omega)\rho(z))} \label{eq:xilambda2}.
\end{eqnarray}
\end{subequations}
We will return to the previous expression later, by the moment we will focus on the Ansatz given by  
\begin{equation}
\xi \lambda(z) = \frac{1 + \alpha H(z)}{3(1+\omega)},
\label{eq:ansatz}
\end{equation}
where $\alpha$ is a constant. It is worthy to mention that with this choice for the $\xi \lambda$ term was possible to obtain singular and non singular cosmological solutions in Rastall theory \cite{epjc1}. If we choose $\alpha = 1/H_{0}$, being $H_{0}$ the Hubble constant, we can write the Eq. (\ref{eq:delta}) as follows
\begin{equation}
\Delta(z) = \frac{3(1+\omega)}{4}\left(\frac{E(z)}{1+E(z)-3(1+\omega)/4}\right),
\label{eq:delta2}
\end{equation}
with $E(z) := H(z)/H_{0}$, which is the normalized Hubble parameter. As we will see below, this form of the $\Delta$ function allows us to find an explicit solution for the normalized Hubble parameter.

%%%%%%%%%%%%%%%%%%%%%%%%
\subsection{HOLOGRAPHIC APPROACH}
%%%%%%%%%%%%%%%%%%%%%%%%

If we consider the definition of the deceleration parameter, $1+q = -\dot{H}/H^{2}$, we can write
\begin{equation}
\rho_{R}(z) = 6\lambda(z)(1-q(z))H^{2}(z),
\label{eq:resembles}
\end{equation}  
note that previous equation resembles the conventional formula of a holographic dark energy model when the Hubble scale is considered and the factor commonly named as $c^{2}$ term, it is not a constant value \cite{radicella, us}, in general we can have $\rho = 3c^{2}(z)H^{2}(z)$, this differs from the Li model in which the $c^{2}$ term is simply a constant value given by $3c^{2}$ \cite{li}. In our case, from Eq. (\ref{eq:resembles}) we can establish that $c^{2}(z) = 2\lambda(z)(1-q(z))$. In order to provide an adequate description of the current status of the universe, the $c^{2}$ term it is expected to be a slowly varying function in the interval $0 < c^{2}(z) < 1$, which according to the value taken by this function, the holographic model could provide a cosmological constant cosmic expansion or an eternal expansion evolution. In general grounds, the $c^{2}$ term depends on the characteristic length assumed by the holographic model. Using the interval in which lies the $c^{2}$ term, we can constraint the value of the Rastall parameter in terms of $q(z)$
\begin{equation}
0 < \lambda(z) < \frac{1}{2[1-q(z)]}.
\label{eq:Rparameter}
\end{equation}    
According to the $\Lambda$CDM model, the deceleration parameter takes the form \cite{wmap}
\begin{equation}
q(z) = -1 + \frac{3}{2\left[1+\frac{\Omega_{\Lambda,0}}{\Omega_{m,0}}(1+z)^{-3}\right]},
\label{eq:cdm}
\end{equation}
where the normalization condition, $\Omega_{\Lambda,0} + \Omega_{m,0} = 1$, must be satisfied. The subscript 0 in the cosmological quantities denotes evaluation at present time, i.e, $z=0$. The recent results reported by the Planck Collaboration \cite{planck2018} reveal that the matter density parameter, $\Omega_{m,0}$, takes the value $\Omega_{m,0} = 0.315 \pm 0.007$, then, for this range of values of $\Omega_{m,0}$, the deceleration parameter given in Eq. (\ref{eq:cdm}) at present time it is restricted to the interval $-0.538 \leq q_{0} \leq -0.517$. From the holographic point of view, at present time the Rastall parameter must be located in the intervals
\begin{align}
& 0 < \lambda_{+,0} < 0.3295, \label{eq:lamplus} \\
& 0 < \lambda_{-,0} < 0.3251. \label{eq:lamminus}
\end{align}
Following the standard definition of the cosmological density parameters, we have
\begin{equation}
\Omega_{R} = \frac{\rho_{R}}{3H^{2}_{0}}.
\label{eq:parameter}
\end{equation}
From the previous equation and the expression given in (\ref{eq:resembles}), we can obtain at present time
\begin{equation}
\lambda_{0} = \frac{\Omega_{R,0}}{2(1-q_{0})}.
\end{equation}
If we assume the region given by $0.678 \leq \Omega_{R,0} \leq 0.692$ for the Rastall density parameter (\ref{eq:parameter}) at present time and the values given previously for the deceleration parameter, we can establish that $0.2234 \leq \lambda_{0} \leq 0.2249$. It is worthy to mention that the obtained interval for $\lambda_{0}$ under the holographic assumption contains the one obtained if we consider that the corresponding Rastall density parameter ($\Omega_{R,0}$) takes values that characterize a cosmological constant dark energy model. According to the $\Lambda$CDM model, the Eq. (\ref{eq:delta}) can be written as follows
\begin{equation}
\Delta(z) = \frac{3}{4}\frac{\left(\xi - \frac{1}{3\lambda(z)} \right)}{\left(\xi - \frac{1}{4\lambda(z)} \right)}.
\end{equation}
If we evaluate $\Delta(z)$ at present time we can observe that its sign depends only on the value of the $\xi$ parameter (if we consider the values obtained before for $\lambda_{0}$). For $\Delta(z) = 1$ the conservation equation (\ref{eq:continuity}) it is fulfilled.\\

$\bullet$ {\bf matter sector:}\\
 Finally, taking into account the Eq. (\ref{eq:accel}) together with the definition of the deceleration parameter in the Eq. (\ref{eq:xilambda1}), we have
\begin{equation}
\rho(z) = 3\underbrace{\left[\frac{1-2\xi \lambda(z)(1-q(z))}{\xi}\right]}_{c^{2}(z)}H^{2},
\end{equation} 
repeating the procedure done in the previous case, we can obtain the following condition if we consider the Eq. (\ref{eq:Rparameter})
\begin{equation}
1 < \frac{1}{\xi} < 2.
\end{equation}
On the other hand, if we consider the Eqs. (\ref{eq:fried}) and (\ref{eq:accel}), we can write the energy density as expected under a holographic description, i.e., $\rho(z) = 3c^{2}(z)H^{2}(z)$, where $c ^{2}(z)$ now has the form
\begin{equation}
c^{2}(z) = 2\lambda(z)\left(\frac{1-q^{2}(z)}{(1+3\omega)/2-q(z)} \right).
\end{equation}
By evaluating at present time the energy density, we obtain the standard definition of the cosmological density parameter, $\rho_{0} = 3 c^{2}_{0} H^{2}_{0} \rightarrow \Omega_{m,0} = c^{2}_{0}$. For $\omega =0$, i.e., the matter sector behaves as dark matter and considering again the values given previously for the deceleration parameter together with $\Omega_{m,0} = 0.315 \pm 0.007$, we obtain the same interval obtained before for $\lambda_{0}$ given by $[0.2234, 0.2249]$.\\

To conclude this section we would like to emphasize what has been done previously, as first step the density $\rho_{R}$ was written in the form of the conventional formula of the holographic approach, $\rho_{R}(z) = 3c^{2}(z)H^{2}(z)$, then from this expression some intervals of validity were obtained for the Rastall parameter $\lambda$. Secondly, we have constructed the density parameter $\Omega_{R}$ associated to the density $\rho_{R}$, later we inserted it into the holographic form of the density, from this last step we found again a value for the Rastall parameter at present time but now assuming that $\Omega_{R}$ corresponds to the cosmological constant dark energy. Finally, by focusing on the matter sector and the conventional holographic formula, we found again the interval of validity for the Rastall parameter at present time but considering that the matter sector behaves as dark matter.\\ 

As can be seen from the previous results, as we approach to the $\Lambda$CDM model, i.e., $\Omega_{R,0} = \Omega_{\Lambda,0}$ or $\omega = 0$, the values for the Rastall parameter are the same and are within the intervals obtained from the holographic point of view. Some acceptable values of certain cosmological parameters constrained with the use of observational data were used, such as the deceleration parameter \cite{wmap} and the density parameters associated to the dark energy and matter sectors \cite{planck2018, planck}.

%%%%%%%%%%%%%%%%%%%%%%
\section{Hubble parameter}
\label{sec:hubble}
%%%%%%%%%%%%%%%%%%%%%%%

\begin{figure}[htbp!]
\centering
\includegraphics[scale=0.65]{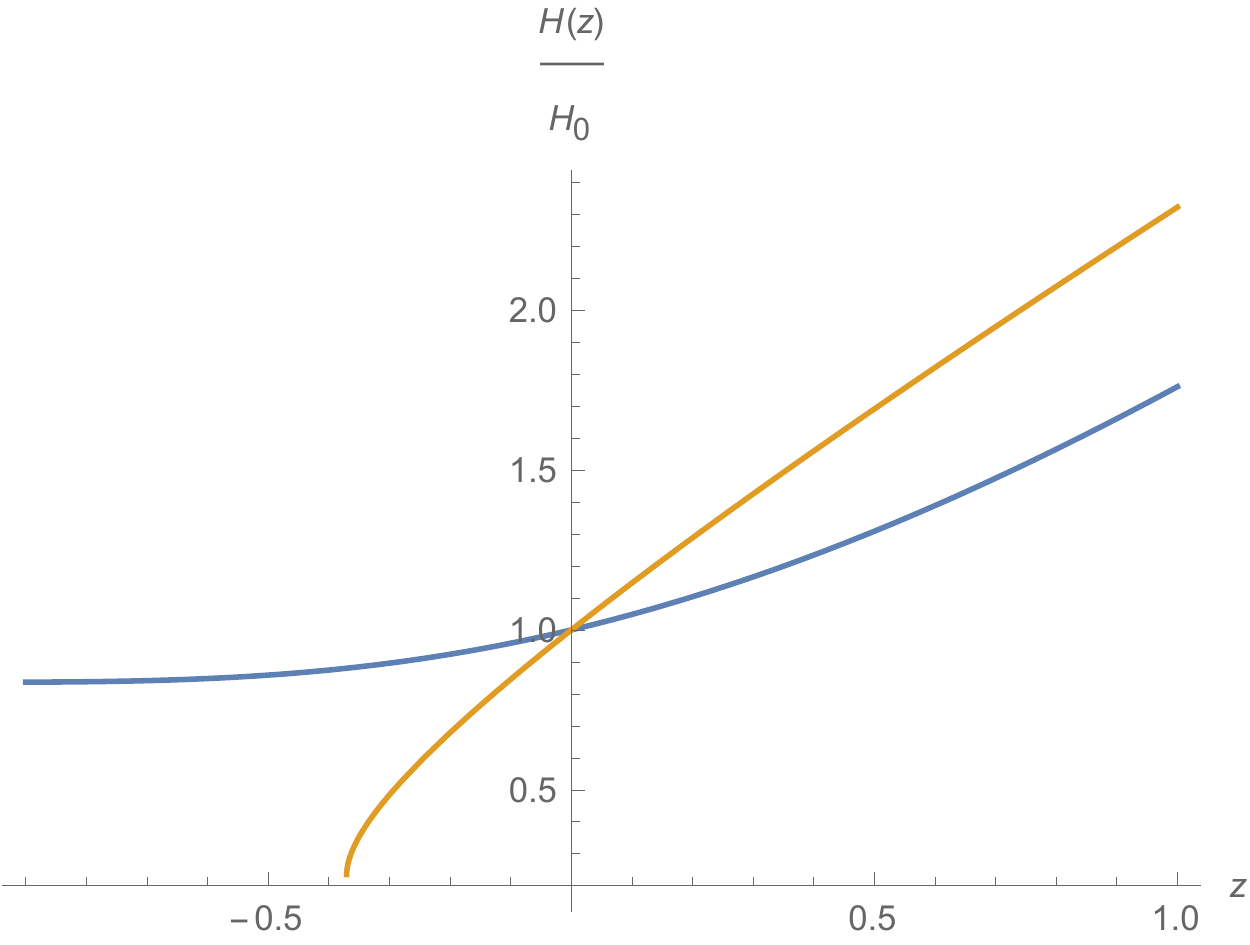}
\caption{Normalized Hubble parameter, the blue line represents the $\Lambda$CDM model with $\Omega_{\Lambda} = 0.692$ and $\Omega_{m,0} = 0.308$.} 
\label{fig:hubble}
\end{figure}

By integrating the Eq. (\ref{eq:continuity}) we can write
\begin{equation}
\rho(z) = \rho_{0}\left(\frac{\Delta_{0}}{\Delta(z)}\right)(1+z) ^{3(1+\omega)/\Delta(z)},
\label{eq:dens}
\end{equation}
where $\rho_{0}$ and $\Delta_{0}$ are constant values. Additionally, using the Eqs. (\ref{eq:fried}), (\ref{eq:accel}) and the definition of the density $\rho_{R}$, we obtain
\begin{equation}
3H^{2}(z) = \xi \Delta(z) \rho(z). 
\end{equation}
This last expression can be rewritten as follows
\begin{equation}
H^{2}(z) = H^{2}_{0}(1+z)^{3(1+\omega)/\Delta(z)},
\label{eq:hubblephan}
\end{equation}
where we have chosen conveniently $\xi = 3H ^{2}_{0}/(\rho_{0}\Delta_{0})$ and Eq. (\ref{eq:dens}). If we substitute the expression given by Eq. (\ref{eq:delta2}) in the above equation we can solve for the normalized Hubble parameter, obtaining
\begin{equation}
E(z) = \frac{\left[ \ln (1+z)^{2}-\ln (1+z)^{3(1+\omega)/2} \right]}{W\left(\frac{[4-3(1+\omega)]\ln (1+z)}{2(1+z)^{2}}\right)},
\end{equation}
where $W(z)$ is the Lambert function. In the Fig. (\ref{fig:hubble}) we show the behaviour of the normalized Hubble parameter when we consider $\omega = 0$ and we perform a comparison with the $\Lambda$CDM model (blue line),  as shown in the plot at present time $E(z) = 1$ in both cases, as expected. Some comments are in order for the Rastall model: towards the future the value of the normalized Hubble parameter decreases faster than in the $\Lambda$CDM model and reaches a specific value (not zero), this behaviour it is inherited from the Lambert function, which has a {\it stationary point}, i.e., the model will have a positive Hubble parameter in the interval $-0.4 \lesssim z < \infty$, we must remember that also in the $\Lambda$CDM model the Hubble parameter has a bounded value in the far future, $E(z \rightarrow -1) \rightarrow \sqrt{\Omega_{\Lambda}}$. On the other hand, the main difference between the Rastall theory and the $\Lambda$CDM model is given at the past, the growth of the Hubble parameter in the latter case it is unbounded and given by $E(z \rightarrow \infty) \rightarrow (1+z)^{3/2}\sqrt{\Omega_{m,0}}$, as it is well known the Lambert function it is unbounded for $z \rightarrow \infty$, therefore at the past the Hubble parameter of the Rastall model exhibits an asymptotic behaviour.  
%%%%%%%%%%%%%%%%%%%%%%%%%%%%%%%%%%%%%%%%%%%%%%%%%%%%%
%%%%%%%%%%%%%%%%%%%%%%%%%%%%%%%%%%%%%%%%%%%%%%%%%%%%

%%%%%%%%%%%%%%%%%%%%%%%%
\subsection{ADMITTING PHANTOM BEHAVIOUR}
%%%%%%%%%%%%%%%%%%%%%%%%

If we assume that at some given value of the redshift namely, $\bar{z}$, a big rip singularity could take place, we must consider that $-1 < \bar{z} < 0$. For this specific value of the redshift we must have for the Hubble parameter, $H(z\rightarrow \bar{z}) \rightarrow \infty$, it is worthy to mention that this is not the unique criterion to determine if the singularity corresponds to a genuine big rip, in Ref. \cite{class} we can find a general classification of the future singularities according to the behaviour of some cosmological quantities (and their derivatives) such as the scale factor, the density and the pressure of the fluid. As observed in Eq. (\ref{eq:hubblephan}), the aforementioned condition for the Hubble parameter can be achieved if $\Delta(z \rightarrow \bar{z})\rightarrow 0$. If we consider the function $\Delta(z)$ as given in Eq. (\ref{eq:delta}) we get, $\Delta(\bar{z}) = 0$ if $\xi \lambda(\bar{z}) = 1/3(1+\omega)$. We must note that the phantom behaviour in order to be admitted depends on the election of the $\xi \lambda(z)$ term. From the definition of the redshift, the continuity equation given in (\ref{eq:continuity}) can be expressed as follows
\begin{equation}
\frac{d\rho(z)}{dz} = \frac{1}{\Delta(z)}\left[3\left(\frac{1+\omega}{1+z}\right)-\frac{d\Delta(z)}{dz}\right]\rho(z),
\label{eq:contz}
\end{equation}
the phantom regime is characterized by $\omega < -1$, therefore
\begin{equation}
\frac{d\rho(z)}{dz} = -\frac{1}{\Delta(z)}\left[3\left(\frac{\left|\omega \right|-1}{1+z}\right)+\frac{d\Delta(z)}{dz}\right]\rho(z),
\label{eq:contphan}
\end{equation}
according to \cite{class} for a big rip singularity we must have $\rho(z \rightarrow \bar{z})\rightarrow \infty$, this implies that $(d\rho(z)/dz)_{z \rightarrow \bar{z}}\rightarrow - \infty$, then from the continuity equation (\ref{eq:contphan}) we can have a benchmark to construct the $\Delta(z)$ function by the appropriate election of the $\xi \lambda(z)$ term if we want to allow a phantom behaviour, i.e., the Rastall parameter plays an important role in order to have a big rip cosmology in the model. 

%%%%%%%%%%%%%%%%%%%%%%
\section{Second law and temperature} 
\label{sec:second}
%%%%%%%%%%%%%%%%%%%%%
The continuity equation (\ref{eq:continuity}) can be written also as
\begin{equation}
\dot{\rho}+3H(1+\omega)\rho = \frac{d}{dt}\left[(1-\Delta)\rho \right]
\label{eq:rhosecond}
 \end{equation}
and according to the second law of thermodynamics \cite{callen}, we can write
\begin{equation}
TdS = d(\rho V) +pdV,
\label{eq:second}
\end{equation}
where $T$ is the temperature of the system, $S$ the entropy, $\rho V$ is the internal energy and $V$ is the Hubble volume given by $V(a) = V_{0}(a/a_{0})^{3}$ such that $dV = 3V_{0}(a/a_{0})^{2}d(a/a_{0})$ then $dV/V = 3(a/a_{0})^{-1}d(a/a_{0}) = 3Hdt$. By taking the time derivative of the Eq. (\ref{eq:second}) and considering a barotropic equation of state together with equation (\ref{eq:rhosecond}), one gets
\begin{equation}
T\frac{dS}{dt} =  V\frac{d}{dt}\left[(1-\Delta)\rho \right] \ \ \ \mbox{or} \ \ \ T\frac{dS}{dz} =  V\frac{d}{dz}\left[(1-\Delta)\rho \right]. 
\end{equation}
If we consider the $\Delta$ function given in Eq. (\ref{eq:delta}), the previous expression takes the form
\begin{equation}
T\frac{dS}{dz} = V\frac{d}{dz}\left[\left\lbrace \frac{(1-3\omega)}{4}\frac{\xi \lambda(z)}{\xi \lambda(z)-1/4} \right\rbrace \rho(z) \right].
\label{eq:second1}
\end{equation}
It is worthy to mention that in the Rastall model if we consider $T \neq 0$, the adiabatic expansion $(S=\mbox{constant})$ it is obtained only with the value $\omega = 1/3$. This result differs from the standard cosmology where the expansion it is adiabatic for a single fluid.\\ 

In order to evaluate the temperature we can write the Eq. (\ref{eq:rhosecond}) in a convenient form as follows
\begin{equation}
\dot{\rho} + 3H(1+\omega_{eff})\rho = 0,
\end{equation}
where we have defined $\omega_{eff}:= \omega - (1/3H\rho)d[(1-\Delta)\rho]/dt = \omega + [(1+z)/3\rho]d[(1-\Delta)\rho]/dz$. According to \cite{maartens}, the evolution equation for the temperature is given by
\begin{equation}
\frac{\dot{T}}{T} = -3H\frac{\partial p}{\partial \rho} \ \ \ \Rightarrow \ \ \ \frac{1}{T}\frac{dT}{dz} = 3\omega_{eff}(1+z)^{-1},
\end{equation}
where $\rho$ and $p$ are related through the parameter state $\omega_{eff}$. Solving the previous expression for the temperature we obtain
\begin{align}
& T(z) = T_{0} \exp \left\lbrace 3\int^{z}_{0} \left[\omega+\frac{1+z'}{3\rho}\frac{d[(1-\Delta(z'))\rho]}{dz'} \right]\frac{dz'}{1+z'}\right\rbrace, \nonumber \\
& = T_{0}(1+z)^{3\omega}\exp \left\lbrace \int^{z}_{0}(1-\Delta(z'))d \ln \left[(1-\Delta(z'))\rho(z') \right] \right\rbrace,
\label{eq:temp}
\end{align}
where $T_{0}$ is a constant value, as can be observed from the previous result, an explicit expression for the temperature can be obtained only when we choose an adequate $\xi \lambda(z)$ term to construct the $\Delta(z)$ function, then from the thermodynamics perspective we can have an additional criterion to build the $\xi \lambda(z)$ term such that could provide a positive temperature. If we consider the value $\omega = 1/3$ in the Eq. (\ref{eq:delta}) we have $\Delta(z) = 1$, therefore we have for the temperature, $T(z) = T_{0}(1+z)$, which is the expected result in an adiabatic expanding universe within the single fluid description. By evaluating at present time the latter expression expression for the temperature we obtain the constant value, $T_{0}$, that can be associated to the temperature of the CMB spectra. On the other hand, for $\Delta \neq 1$, we will have non adiabatic expansion, the temperature in this case will be given as a function of the redshift (also for the case $\omega = 0$). However, some results show that a non adiabatic expansion is more consistent from the thermodynamics perspective than an adiabatic expansion \cite{carde, pavon}. In general, the universe cools down as evolves towards the future, as can be seen from Eq. (\ref{eq:temp}).\\

On the other hand, to be in agreement with the second law of thermodynamics we must have $dS/dt > 0$ which in terms of the redshift can be written as $dS/dz < 0$. If we consider the Eqs. (\ref{eq:delta}), (\ref{eq:contz}) and (\ref{eq:second1}), we can write
\begin{equation}
\frac{dS}{dz} = \frac{1-3\omega}{1+z}\left\lbrace \frac{1-\frac{(1+z)}{3(1+\omega)(4\xi \lambda(z)-1)}\frac{d\ln (\xi \lambda(z))}{dz}}{1-\frac{1}{3\xi \lambda(z)(1+\omega)}}\right\rbrace \rho(z).
\label{eq:derivative}
\end{equation}
In order to visualize the behaviour of the previous expression we will consider $\xi \lambda(z) \rightarrow 1$ together with $-1 < z < \infty$, in this case we will have that if $\rho(z) > 0$ and $\omega > 1/3$ then $dS/dz < 0$, as we approach to the general relativity case, $\xi \lambda(z) \rightarrow 0$, the term within the braces in Eq. (\ref{eq:derivative}) tends to $-1$, therefore the sign of the first derivative of the entropy will depend again on the election of the parameter state and the density of the fluid as follows, for $\omega < 1/3$ the density must be positive. As in the previous results, for a definite answer about the behaviour of the equation (\ref{eq:derivative}), we strongly depend on the election of the $\xi \lambda(z)$ term. For thermodynamics consistency, any cosmological model must obey simultaneously two conditions for the entropy: $dS/dt > 0$ and the convexity condition given by $d^{2}S/dt^{2} < 0$ \cite{lepe, manuel}, for the latter condition we can write in terms of the redshift
\begin{equation}
0 < \frac{d^{2}S}{dz^{2}} < - \left(\frac{1}{1+z}+\frac{d \ln H(z)}{dz} \right)\frac{dS}{dz},
\end{equation}    
it is important to point out that the r.h.s. of the previous inequality it is a positive quantity since $dS/dz < 0$ (second law of thermodynamics). Using the expression (\ref{eq:delta}) and the Hubble parameter written in Eq. (\ref{eq:hubblephan}), one gets
\begin{widetext}
\begin{eqnarray}
0 < \frac{d^{2}S}{dz^{2}} &<& (1-3\omega)\xi \lambda(z)\left[\left\lbrace 1 - 3(1+\omega)\ln (1+z)\right\rbrace (1+z)\frac{d \ln (\xi \lambda(z))}{dz}-3(1+\omega)(4\xi \lambda(z)-1)\right]\nonumber \times \\
&\times & \frac{\left[3\omega + 3(1+\omega)\xi \lambda(z) \left\lbrace 15\xi \lambda(z)(1+\omega)-3(3+\omega) \right\rbrace + 4 \right]}{(1+z)^{2}(4\xi \lambda(z)-1)[3\xi \lambda(z)(1+\omega)-1]^{3}}\rho(z).
\label{eq:secondderivative}
\end{eqnarray}
\end{widetext}
By considering again the case $\xi \lambda(z) \rightarrow 1$ and $-1 < z < \infty$, we can observe that the factor of $\rho(z)$ at the r.h.s. of the previous expression is negative, therefore for $\omega > 1/3$ and $\rho(z) > 0$ we will have $d^{2}S/dz^{2} > 0$, note that for this election of values we have $dS/dz < 0$, then the model it is consistent from the thermodynamics point of view. On the other hand, as we approach to the Einstein gravity, $\xi \lambda(z) \rightarrow 0$, the factor of $\rho(z)$ in the Eq. (\ref{eq:secondderivative}) is always positive, therefore for $\omega < 1/3$ the positivity it is guaranteed with $\rho(z) > 0$, these values also lead to the condition $dS/dz < 0$ in the Einstein gravity limit discussed previously.\\ 

As commented before, once we choose an appropriate $\xi \lambda(z)$ term, we could evaluate these conditions explicitly, however in general grounds we can see that the Rastall gravity it is consistent from the thermodynamics perspective. In Ref. \cite{epjc2} was found that the thermodynamics consistency at the apparent horizon using the Rastall model can also be guaranteed if some corrections to the entropy are included, however cases like logarithmic or power law corrections lead to instability of thermodynamics equilibrium.   

%%%%%%%%%%%%%%%%%%%%%%
\section{Creation of particles} 
\label{sec:creation}
%%%%%%%%%%%%%%%%%%%%%

The matter production in spacetime has been studied from different points of view. However, several models can present some inconsistencies at quantum level if one desires to apply some aspects of the quantum field theory \cite{maartens2}. A possible way to solve these issues is given by complementing the Einstein's equations with the particle production effects. Following this line of reasoning, in Ref. \cite{prigogine} was found within the context of thermodynamics of open systems that the production of matter can lead to a growing behaviour for the entropy, nonetheless all processes must be considered as irreversible. In this scheme the production of particles can be understood only by re-interpreting the resulting energy-momentum tensor, where an extra term emerges playing the role of a negative pressure. In the literature can be found that the production/annihilation of particles at cosmological level is due to the presence of dissipative effects \cite{zeldovich}, such effects lead to deviations from the local equilibrium pressure \cite{eckart, stewart, zimdahl1, zimdahl2}. Adopting a thermodynamic perspective as given in \cite{prigogine, lima}, from the Gibbs relation we have
\begin{equation}
nTdS = d\rho - (\rho + p)\frac{dn}{n},
\label{eq:gibbs}
\end{equation}     
where $n$ is the particle number density, $n=N/V$, being $N$ the number of particles in the observable universe, since we can have production/annihilation of particles at some rate, $\Gamma$, we have non-conservation of particle number, therefore
\begin{equation}
\dot{n} + 3Hn = n\Gamma,
\label{eq:noncons}
\end{equation}
where $\Gamma$ acts as a source or sink of particles if $\Gamma > 0$ or $\Gamma < 0$, respectively. As we will see below, $\Gamma$ contributes to the entropy production. By taking the time derivative of Eq. (\ref{eq:gibbs}) we have
\begin{align}
nT\frac{dS}{dt} & = \dot{\rho} - (\rho + p)\frac{\dot{n}}{n},\nonumber \\
& = \dot{\rho} + 3H\rho(1+\omega)\left[1-\frac{\Gamma}{3H} \right],\nonumber\\
& = -\left[\frac{d \ln \Delta}{dt}+3H(1+\omega)\left(\frac{1}{\Delta}-1+\frac{\Gamma}{3H} \right) \right]\rho,
\end{align}
where the expressions (\ref{eq:continuity}), (\ref{eq:delta}) and (\ref{eq:noncons}) were used together with the barotropic equation of state. It is worthy to mention that if we consider the special case $\omega = 1/3$, we obtain $\Delta = 1$, in the previous section these values led to adiabatic expansion, however once we introduce the production/annihilation of particles, the adiabatic expansion is not longer available since $\Gamma \neq 0$ for consistency, i.e., $dS/dt \propto \Gamma \rho$. Therefore, in this regime we do not share the idea discussed in Ref. \cite{epjc1}, where the adiabatic expansion was claimed in order to explore the matter production in Rastall gravity. From the previous result, we can observe that the production/annihilation rate of particles must obey the following condition
\begin{equation}
\Gamma < -\frac{1}{1+\omega}\left[\frac{d \ln \Delta}{dt}+ 3H(1+\omega)\left(\frac{1-\Delta}{\Delta}\right) \right],
\end{equation}
in order to be in agreement with the second law of thermodynamics, $dS/dt > 0$. If we take the definition of the $\Delta$ function given in Eq. (\ref{eq:delta}), the previous condition for $\Gamma$ is simply given as follows
\begin{equation}
\Gamma(t) < -(1-3\omega)\frac{\left[\xi \dot{\lambda}(t)-3\xi \lambda(t)H(t)(1+\omega)\left\lbrace 1-4\xi \lambda(t)\right\rbrace \right]}{(1+\omega)[1-4\xi \lambda(t)][1-3\xi \lambda(t)(1+\omega)]},
\end{equation}
taking into account the Ansatz (\ref{eq:ansatz}) with $\alpha = 1/H_{0}$ and the definition of the deceleration parameter, we can write
\begin{widetext}
\begin{equation}
\frac{\Gamma(t)}{H_{0}} < (1-3\omega)\frac{[1+q(t)+36(\xi \lambda(t))^{2}(1+\omega)^{2}-3\xi \lambda(t)(1+\omega)(4+3\omega +q(t))]}{3(1+\omega)^{2}[1-4\xi \lambda(t)]}.
\label{eq:quotient}
\end{equation}
\end{widetext}

%%%%%%%%%%%%%%%%%%%%%%%%%%%%%%%%%%%%%%%%%%%%%%%%5
\onecolumngrid

\begin{figure}[htbp!]
\centering
\includegraphics[width=7.5cm,height=5.5cm]{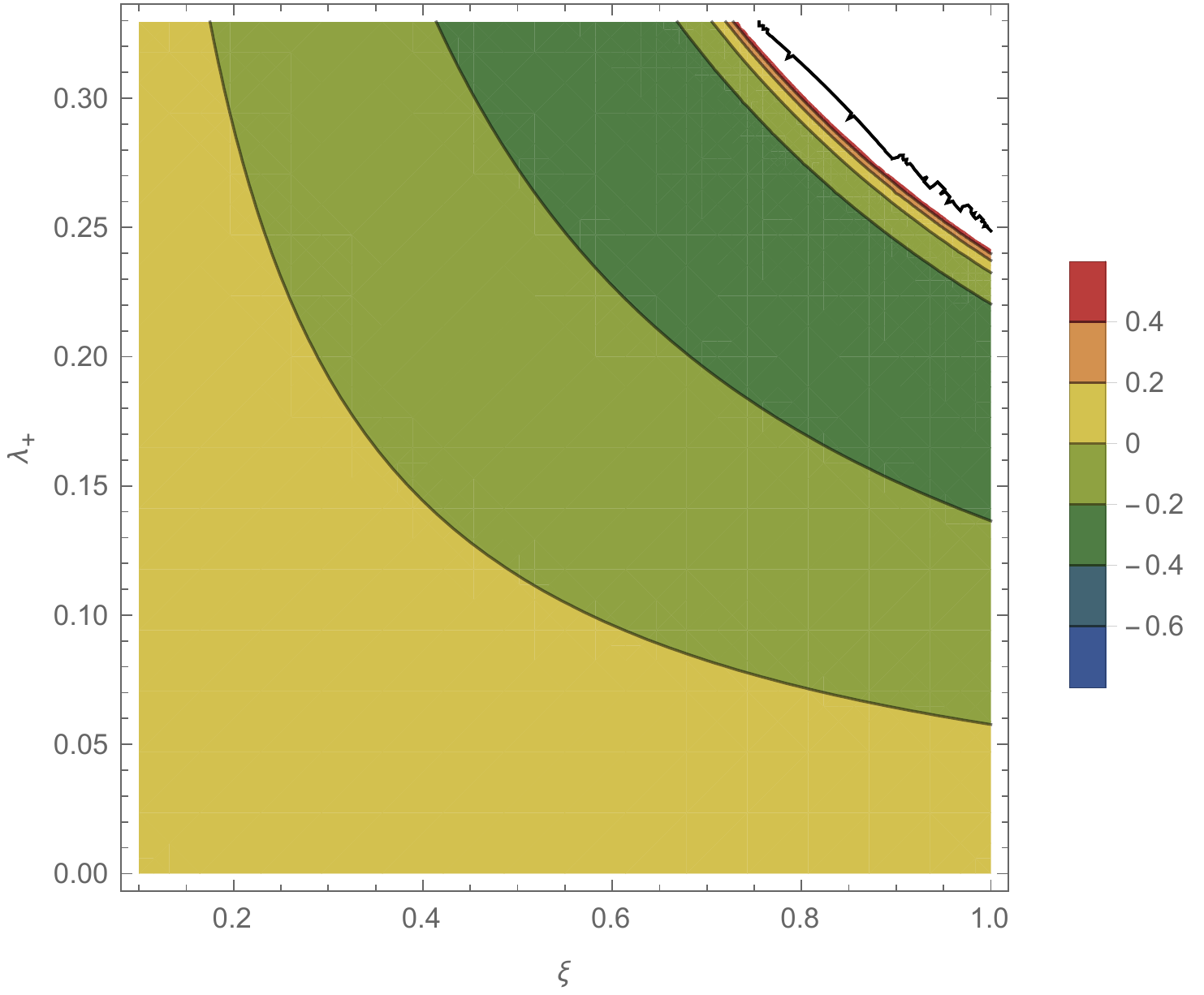}
\includegraphics[width=7.5cm,height=5.5cm]{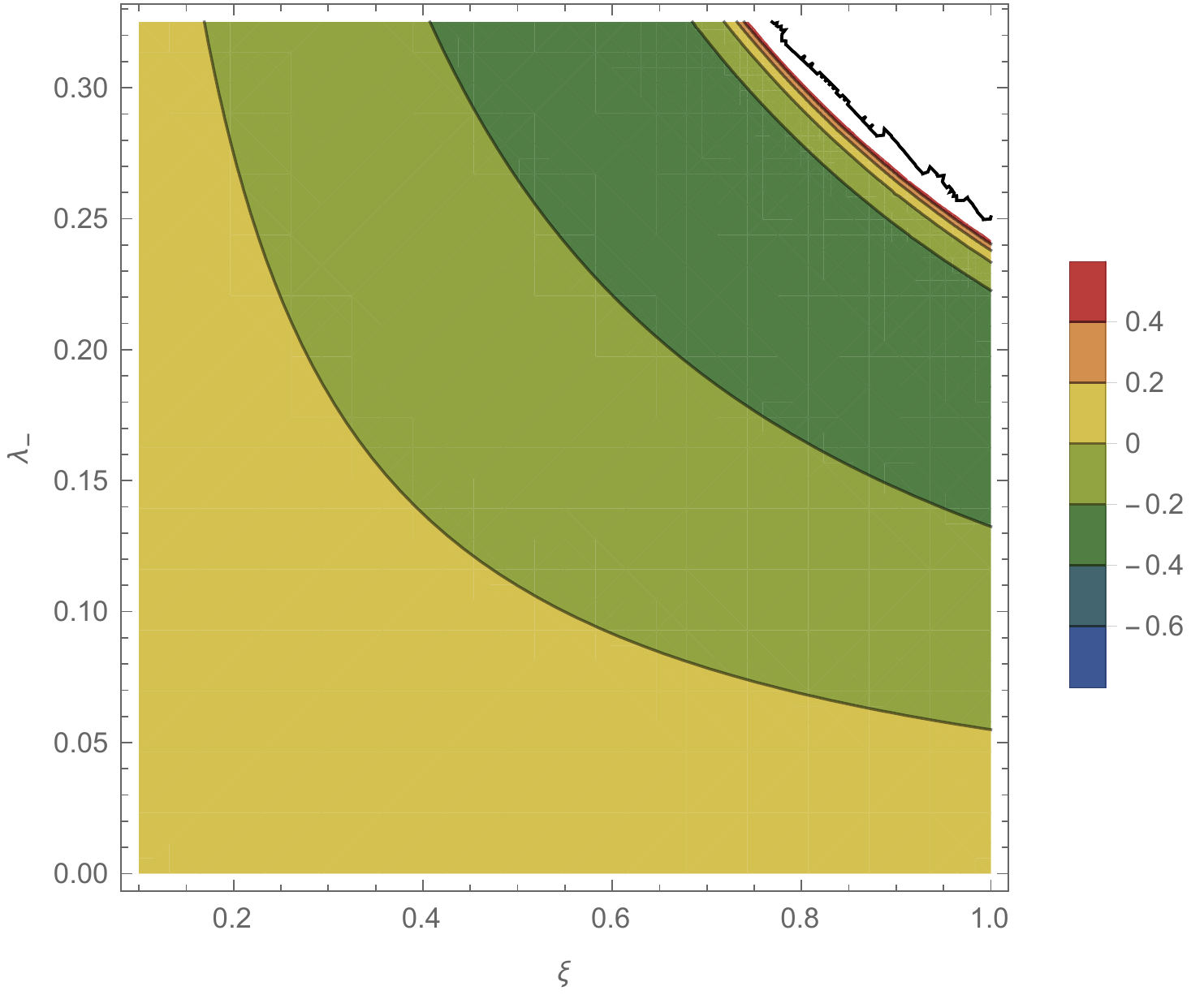}
\caption{Behaviour of the quotient $\Gamma_{0}/H_{0}$. The value $\omega = 0$ was considered in both panels together with $-0.538 \leq q_{0} \leq -0.517$ for the deceleration parameter.} 
\label{fig:productionrate}
\end{figure}

\twocolumngrid
%%%%%%%%%%%%%%%%%%%%%%%%%%%%%%%%%%%%%%%%%%%% 

From the previous expression we can determine the production/annihilation rate of particles at present time, for this purpose we will use the values obtained previously for the Rastall parameter within the holographic scheme, $\lambda_{\pm,0}$, given in Eqs. (\ref{eq:lamplus}) and (\ref{eq:lamminus}).\\
 
In Fig. (\ref{fig:productionrate}) we show the quotient, $\Gamma_{0}/H_{0}$, given as a function of the parameters $(\xi, \lambda_{\pm,0})$. As can be seen from the plots, depending on the values of $(\xi, \lambda_{\pm,0})$, the production or annihilation of particles at present time can be possible, we restrict ourselves to the interval $0 < \xi < 1$, it is worthy to mention that in the limit $\xi \rightarrow 1$ and $\lambda \rightarrow 0$, we recover the general relativity case.  Then, always that the value of the quotient (\ref{eq:quotient}) be less than the values bounded by the regions shown in both panels of Fig. (\ref{fig:productionrate}), the model will be in agreement with the second law of thermodynamics.  

%%%%%%%%%%%%%%%%%%%%%%
\section{Chemical potential} 
\label{sec:chemical}
%%%%%%%%%%%%%%%%%%%%%

In this section we will explore the effects of introducing chemical potential in the Rastall model. In the context of standard cosmology, several works have revealed that the sign of the chemical potential has incidence on the value of the parameter state for dark energy, i.e., for a positive chemical potential the parameter state takes values greater than $-1$ and for a negative chemical potential the phantom regime is allowed, however in this regime some inconsistencies appear leading to negative temperature and positive entropy or vice versa, see for instance the Refs. \cite{lima1, lima2, saridakis}. Nonetheless, in Ref. \cite{odintsovdiss} was found that if a dissipative cosmology is considered in the context of irreversible thermodynamics, the problem of negative entropy or negative temperature can be solved since the dissipative effects allow the construction of a negative chemical potential.\\

The chemical potential, $\mu$, it is introduced in the Gibbs relation (\ref{eq:gibbs}) as follows
\begin{equation}
n\left(T\frac{dS}{dT} + \mu \dot{n}\right) = \dot{\rho} - (\rho + p)\frac{\dot{n}}{n},
\end{equation}   
following a similar procedure as in the previous section we can write
\begin{equation}
nT\frac{dS}{dT} = \dot{\rho} + 3H(1+\omega)\left(1-\frac{\Gamma}{3H} \right)\left\lbrace 1+\frac{\mu n^{2}}{(1+\omega)\rho}\right\rbrace \rho.
\label{eq:gibbschemical}
\end{equation}
As observed in the previous equation, now the production rate of particles and the chemical potential contribute to the entropy production. In order to discuss the adiabatic case for the entropy we define 
\begin{equation}
1+\omega_{eff}:= (1+\omega)\left(1-\frac{\Gamma}{3H} \right)\left\lbrace 1+\frac{\mu n^{2}}{(1+\omega)\rho} \right\rbrace,
\end{equation}
therefore the continuity equation, $\dot{\rho} + 3H(1+\omega_{eff})\rho = 0$, could assure this condition for the entropy, i.e, $S = \mbox{constant}$. For the specific case $\omega = 0$ one gets
\begin{equation}
\omega_{eff} = -1 + \left(1-\frac{\Gamma}{3H} \right)\left\lbrace 1+\frac{\mu n^{2}}{\rho} \right\rbrace,
\end{equation}
therefore for a positive chemical potential some cases can be obtained for this effective parameter state:
\begin{itemize}
\item 
$\omega_{eff} = 0$ if $\frac{\Gamma}{3H} =1-\left\lbrace 1+\frac{\mu n^{2}}{\rho} \right\rbrace^{-1}$,
\item
$\omega_{eff} < -1$ if $\frac{\Gamma}{3H} > 1$,
\item
$\omega_{eff} = -1$ if $\Gamma = 3H$.
\end{itemize}
At effective level we could have a phantom regime or a cosmological {\it constant} evolution and this uniquely depends on the values of $\Gamma$ and the chemical potential. However, if we write the explicit form of Eq. (\ref{eq:gibbschemical}) we have
\begin{widetext}
\begin{equation}
nT\frac{dS}{dt} = -\left\lbrace \frac{d \ln \Delta}{dt}+3H(1+\omega)\left[\frac{1}{\Delta} - \left(1-\frac{\Gamma}{3H} \right)\left\lbrace 1+\frac{\mu n^{2}}{(1+\omega)\rho}\right\rbrace \right] \right\rbrace \rho,
\end{equation}
\end{widetext}
as discussed previously, when no other effects such as matter production are considered, the adiabatic case for the entropy is obtained with the value $\omega = 1/3$ which leads to $\Delta = 1$. By considering the aforementioned values for $\omega$ and $\Delta$ in the previous equation, one gets
\begin{equation}
nT\frac{dS}{dt} = -\left\lbrace 4H\left[1 - \left(1-\frac{\Gamma}{3H} \right)\left\lbrace 1+\frac{3\mu n^{2}}{4\rho}\right\rbrace \right] \right\rbrace \rho,
\end{equation}
therefore $S$ is not a constant since $\Gamma, \mu \neq 0$. However, this case is in agreement with the second law of thermodynamics, $dS/dt > 0$, always that
\begin{align}
& \frac{\Gamma}{3H} < 1-\left\lbrace 1+\frac{3\mu n^{2}}{4\rho}\right\rbrace^{-1}, \ \ \ \mbox{with} \ \ \ \mu > 0,\\
& \frac{\Gamma}{3H} < 1-\left\lbrace 1-\frac{3 \left|\mu \right| n^{2}}{4\rho}\right\rbrace^{-1}, \ \ \ \mbox{with} \ \ \ \mu < 0.
\end{align}
In general, the inclusion of the chemical potential does not lead to contradictions at thermodynamics level in the Rastall model.

%%%%%%%%%%%%%%%%%%%%%%
\section{Final remarks} 
\label{sec:elfinal}
%%%%%%%%%%%%%%%%%%%%%

In this work, we focused on general aspects of the Rastall gravity and some of its implications at thermodynamics level.  As it is well known, if we perform a comparison with the Einstein gravity, it results that the Rastall model can be seen as the standard vacuum Einstein model or as Einstein gravity plus a varying cosmological {\it constant}, which it is well known as Rastall parameter, then from this latter point of view, this theory is beyond the standard cosmological model.\\ 

As discussed in our analysis, in order to obtain specific results we must construct an appropriate $\xi \lambda$ term, being $\lambda$ the Rastall parameter and in general this parameter can be a function of the cosmological time or the redshift. An estimation for the $\lambda$ parameter at present time was obtained by means of the holographic approach for the energy density coming from the Rastall model contribution, of course, the values obtained for this parameter depend on the conditions imposed by the holographic approach, being the most important the range of values allowed for the $c^{2}$ term that appears in the conventional holographic formula. The $c^{2}$ term it is constrained to the interval $(0,1)$ to describe an expanding universe at late times. An interesting feature of this procedure is that the Rastall parameter can be written in terms of the deceleration parameter, $q$, which has been estimated with the use of observational data in several dark energy models.\\

In general, there is not a definitive physical criterion to construct the $\xi \lambda$ term, however adopting an Ansatz given in Eq. (\ref{eq:ansatz}) for this term, it is possible to obtain an expression for the Hubble parameter by integrating the continuity equation of this model, for this specific Ansatz we found that the Hubble parameter tends to a bounded value as the universe evolves towards the future, this characteristic is also obtained in the $\Lambda$CDM model. However, the Rastall model reaches this value faster. By inspecting the continuity equation we found that if we desire to allow the phantom regime, we can have a criterion to construct the $\xi \lambda$ term.\\

On the other hand, if we keep the Ansatz philosophy we can find that unlike what happens in standard cosmology for a single fluid description, the condition of adiabatic expansion in the Rastall model depends on the fluid describing the content of the universe, i.e, only the value $\omega = 1/3$ leads to adiabatic expansion. In general grounds, the corresponding temperature of the universe will keep cooling down along the evolution. Likewise, for other possible values of the parameter state, $\omega$, the model is consistent at the thermodynamics level, that is, the second law is fulfilled simultaneously with the condition of convexity for entropy. Therefore, from the results discussed in our work, adopting a thermodynamic perspective could give us a strong criterion to construct the $\xi \lambda$ term which could keep the model well defined for this approach.\\

Besides, once the effects of non conservation of matter are considered, the adiabatic case is no longer available since the production rate of particles contributes to the entropy production, this does not imply that some thermodynamics inconsistencies may arise in the model. In order to know what kind of behaviour we would be observing today, using the value of the Rastall parameter obtained with the holographic description, we can see that we could have creation or annihilation of matter and this entirely depends on the value of the parameter $\xi$. Finally, it is interesting that the inclusion of chemical potential in our description allows us to have at least at effective level, a cosmological constant or even phantom behaviour for the fluid with the advantage of keeping the thermodynamics consistency, we will explore this elsewhere.  

%%%%%%%%%%%%%%%%%%%%%%%
\section*{Acknowledgments}
%%%%%%%%%%%%%%%%%%%%%%%
M. C. work has been supported by S.N.I. (CONACyT-M\'exico). G. M. N. also acknowledges partial support from a SNI-CONACyT scholarship.

\end{document}